\documentclass{jps-cp}
\usepackage{txfonts} %Please comment out this line unless the txfonts package is availabe in your LaTeX system.
\usepackage{bm}
\usepackage{color}

\title{Equation of State for Neutron Stars in the Quark-Meson Coupling Model with the Cloudy Bag}

\author{Tsuyoshi~\textsc{Miyatsu} and Koichi~\textsc{Saito}}

\inst{Department of Physics, Faculty of Science and Technology, Tokyo University of Science, Noda 278-8510, Japan}

\email{tsuyoshi.miyatsu@rs.tus.ac.jp}

\recdate{January 25, 2019}

\abst{Using the quark-meson coupling model with the cloudy bag, we construct the equation of state for neutron stars with hyperons in SU(3) flavor symmetry.
The hyperfine interaction due to the gluon exchange and the pion-cloud effect inside a baryon is taken into account in vacuum and nuclear matter.
We investigate how the quark degrees of freedom and the exchanges of gluon and pion between two quarks affect the properties of nuclear and neutron-star matter.
It is found that the variation of baryon substructure in matter plays an important role in supporting massive neutron stars.}

\kword{equation of state, dense nuclear matter, neutron stars, hyperons}

\begin{document}
\maketitle

\section{Introduction}

Many theoretical discussions on the properties of nuclear matter have been studied so far, where protons and neutrons are generally treated as point-like objects as in Quantum Hadrodynamics (QHD) ~\cite{Serot:1984ey}.
However, it is apparent that baryons are composed of quarks and gluons.
There are a few calculations in which the quark degrees of freedom inside a baryon and the variation of baryon substructure in matter are considered.

Thanks to the recent advances in astrophysical observations, we can get some precise information on the properties of neutron stars.
In particular, the discovery of massive neutron stars~\cite{Demorest:2010bx,Antoniadis:2013pzd} is very useful to construct the equation of state (EoS) for dense nuclear matter.
In addition, the gravitational wave from binary neutron star detected by LIGO and Virgo scientific collaborations~\cite{TheLIGOScientific:2017qsa} is also important to make a constraint on the EoS for neutron stars.

In the present study, we consider the baryon substructure due to quarks using the quark-meson coupling (QMC) model~\cite{Guichon:1987jp,Saito:1994ki,Saito:2005rv}, and construct the EoS for neutron stars with hyperons, which can satisfy with both nuclear properties and astrophysical observations.

\section{Matter Description at the Quark-Mean Field Level}

Using the volume coupling version of the cloudy bag model (CBM)~\cite{Thomas:1982kv}, the hyperfine interaction due to the gluon exchange and the pion-cloud effect inside a baryon can be taken into account.
The mass for baryon ($B$) with the one-gluon and one-pion exchange effects ($\varDelta E_{g}$ and $\varDelta E_{\pi}$) is described by $M_{B}=E_{\rm MIT}+\varDelta E_{g}+\varDelta E_{\pi}$, where $E_{\rm MIT}$ is the bag energy based on the MIT bag model.
The calculation of baryon spectra is well tuned so as to fit the mass differences between $N$, $\Delta$, and $\Omega$ in vacuum~\cite{Nagai:2008ai,Miyatsu:2010zz}.

In order to describe nuclear matter, we need the intermediate attractive and short-range repulsive nuclear forces.
As in the case of QHD, it is achieved by introducing the $\sigma$ and $\omega$ mesons.
In addition, the strange mesons ($\sigma^{\ast}$ and $\phi$) and $\bm{\rho}$ meson are included.
The effect of the baryon substructure is then taken into account through two kind of scalar polarizabilities:
\begin{equation}
  C_{B}(\sigma)=-\frac{1}{g_{\sigma B}}\left(\frac{\partial M_{B}^{\ast}}{\partial\sigma}\right),
  \hspace*{0.5cm}
  C_{B}^{\prime}(\sigma^{\ast})=-\frac{1}{g_{\sigma^{\ast}B}}\left(\frac{\partial M_{B}^{\ast}}{\partial\sigma^{\ast}}\right),
\end{equation}
where the effective baryon mass, $M_{B}^{\ast}$, is calculated by the CBM model, and $g_{\sigma B}$ and $g_{\sigma^{\ast}B}$ are the $\sigma$-$B$ and $\sigma^{\ast}$-$B$ coupling constants, respectively.

\section{Field-Dependent Coupling Constants in the QMC Model}

In the QMC model, it has been known that the scalar polarizabilities decrease linearly as the total baryon density, $\rho_{B}$, increases~\cite{Tsushima:1997cu}.
In the present study, we employ the following, field-dependent, simple parametrizations for them~\cite{Miyatsu:2013yta}:
\begin{equation}
  C_{B}(\sigma)
  = b_{B} \left[1-a_{B}\left(g_{\sigma N}\sigma\right)\right],
  \hspace*{0.5cm}
  C_{B}^{\prime}(\sigma^{\ast})
  = b_{B}^{\prime} \left[1-a_{B}^{\prime}\left(g_{\sigma^{\ast}\Lambda}\sigma^{\ast}\right)\right],
  \label{eq:Csigma2}
\end{equation}
where $g_{\sigma N}$ and $g_{\sigma^{\ast}\Lambda}$ are respectively the $\sigma$-$N$ and $\sigma^{\ast}$-$\Lambda$ coupling constants at zero density, and the parameters $a_{B}$, $b_{B}$, $a_{B}^{\prime}$, and $b_{B}^{\prime}$ are tabulated in Table~\ref{tab:QMC-parameter}.
We here present three cases: (a) the QMC model without gluons and pions, (b) the QMC model with the gluon interaction, denoted by QMCg, and (c) the QMC model with the gluon and pion interactions based on  the chiral quark-meson coupling (CQMC) model~\cite{Nagai:2008ai}.
The effect of the variation of baryon structure at the quark level can be described with the parameters $a_{B}$ and $a_{B}^{\prime}$.
In addition, the extra parameters, $b_{B}$ and $b_{B}^{\prime}$, are necessary to express the effect of hyperfine interaction between two quarks.
%%%%%%%%%%%%%%%%%%%%%%%%%%%%%%%%%%%%%%%%%%%%%%%%%%%%%%%%%%%%%%%%%%%%%%%%%%%%%%%
\begin{table}[t]
\caption{\label{tab:QMC-parameter}
Values of $a_{B}$, $b_{B}$, $a_{B}^{\prime}$ and $b_{B}^{\prime}$ in the QMC, QMCg and CQMC models.
}
\begin{tabular}{l|cccc|cccc|cccc}
\hline\hline
\         & \multicolumn{4}{c}{(a) QMC}
          & \multicolumn{4}{c}{(b) QMCg}
          & \multicolumn{4}{c}{(c) CQMC}                                          \\
\cline{2-5}\cline{6-9}\cline{10-13}
$B$       & $a_{B}$~(fm) & $b_{B}$ & $a_{B}^{\prime}$~(fm) & $b_{B}^{\prime}$
          & $a_{B}$~(fm) & $b_{B}$ & $a_{B}^{\prime}$~(fm) & $b_{B}^{\prime}$
          & $a_{B}$~(fm) & $b_{B}$ & $a_{B}^{\prime}$~(fm) & $b_{B}^{\prime}$ \\
\hline
$N$       & 0.179        & 1.00    & ---                   & ---
          & 0.167        & 1.12    & ---                   & ---
          & 0.118        & 1.04    & ---                   & ---              \\
$\Lambda$ & 0.172        & 1.00    & 0.220                 & 1.00
          & 0.160        & 1.17    & 0.270                 & 1.00
          & 0.122        & 1.09    & 0.290                 & 1.00             \\
$\Sigma$  & 0.177        & 1.00    & 0.223                 & 1.00
          & 0.188        & 1.04    & 0.240                 & 1.18
          & 0.184        & 1.02    & 0.277                 & 1.15             \\
$\Xi$     & 0.166        & 1.00    & 0.215                 & 1.00
          & 0.176        & 1.19    & 0.267                 & 1.05
          & 0.181        & 1.15    & 0.292                 & 1.04             \\
$\Delta$  & 0.196        & 1.00    & ---                   & ---
          & 0.216        & 0.89    & ---                   & ---
          & 0.197        & 0.89    & ---                   & ---              \\
\hline\hline
\end{tabular}
\end{table}
%%%%%%%%%%%%%%%%%%%%%%%%%%%%%%%%%%%%%%%%%%%%%%%%%%%%%%%%%%%%%%%%%%%%%%%%%%%%%%%
Using these linear form, we can obtain the field-dependent coupling constants for the scalar mesons,
\begin{equation}
  g_{\sigma B}(\sigma)
  = g_{\sigma B}b_{B}\left[1-\frac{a_{B}}{2}\left(g_{\sigma N}\sigma\right)\right],
  \hspace*{0.5cm}
  g_{\sigma^{\ast}B}(\sigma^{\ast})
  = g_{\sigma^{\ast}B}b_{B}^{\prime}
  \left[1-\frac{a_{B}^{\prime}}{2}\left(g_{\sigma^{\ast}\Lambda}\sigma^{\ast}\right)\right].
\end{equation}
If we set $a_{B}=0$ and $b_{B}=1$, $g_{\sigma B}(\sigma)$ becomes identical to the $\sigma$-$B$ coupling constant in QHD. This is also true of the coupling $g_{\sigma^{\ast}B}(\sigma^{\ast})$.
% Moreover, the in-medium baryon mass can be written as
% %
% \begin{equation}
%   M_{B}^{\ast}\left(\sigma,\sigma^{\ast}\right)
%   = M_{B} - g_{\sigma B}(\sigma)\sigma - g_{\sigma^{\ast}B}(\sigma^{\ast})\sigma^{\ast},
%   \label{eq:effective-mass-QMC}
% \end{equation}
% %
Thus, the Lagrangian density in the QMC, QMCg, and CQMC models is given by replacing the $\sigma$-$B$ and $\sigma^{\ast}$-$B$ coupling constants in QHD with the field-dependent couplings, $g_{\sigma B}(\sigma)$ and $g_{\sigma^{\ast}B}(\sigma^{\ast})$.

\section{Numerical Results}

As in the case of QHD, the coupling constants are determined so as to reproduce the saturation properties.
We determine the couplings of the vector mesons to the baryons in SU(3) flavor symmetry~\cite{Miyatsu:2013yta}.
For the sake of comparison, we present a calculation based on QHD with the nonlinear (NL) potential, $U_{\rm NL}=\frac{1}{3}g_{2}\sigma^{3}+\frac{1}{4}g_{3}\sigma^{4}$.
The coupling constants, $g_{2}$ and $g_{3}$, are chosen so as to adjust the values of the incompressibility, $K_{0}$, and effective nucleon mass, $M_{N}^{\ast}$, in the QMC model.

%%%%%%%%%%%%%%%%%%%%%%%%%%%%%%%%%%%%%%%%%%%%%%%%%%%%%%%%%%%%%%%%%%%%%%%%%%%%%%%
\begin{table}[t]
\caption{\label{tab:matter-properties}
Properties of symmetric nuclear matter at the saturation density, $\rho_{0}=0.16$ fm$^{-3}$.
% The incompressibility, $K_{0}$, and effective nucleon mass, $M_{N}^{\ast}$, in the QHD+NL model are fixed so as to reproduce those in the QMC model.
The nuclear symmetry energy is also fitted so as to reproduce the empirical data, $E_{\rm sym}(\rho_{0})=32.5$ MeV, in all the cases.
The physical quantities are explained in detail in the text.
}
\begin{tabular}{lcccccccc}
\hline\hline
\      & $M_{N}^{\ast}/M_{N}$ & $K_{0}$ & $J_{0}$ & $L$   & $K_{\rm sym}$ & $K_{\rm asy}$ & $K_{{\rm sat},2}$ \\
Model  &                    \ &   (MeV) &   (MeV) & (MeV) &         (MeV) &         (MeV) &             (MeV) \\
\hline
QHD+NL &                0.800 &     285 &  $-571$ &  88.2 &       $-16.4$ &        $-546$ &            $-369$ \\
QMC    &                0.800 &     285 &  $-459$ &  88.2 &       $-17.1$ &        $-547$ &            $-405$ \\
QMCg   &                0.782 &     294 &  $-407$ &  89.0 &       $-13.3$ &        $-547$ &            $-424$ \\
CQMC   &                0.754 &     309 &  $-320$ &  90.3 &        $-6.3$ &        $-548$ &            $-455$ \\
\hline\hline
\end{tabular}
\end{table}
%%%%%%%%%%%%%%%%%%%%%%%%%%%%%%%%%%%%%%%%%%%%%%%%%%%%%%%%%%%%%%%%%%%%%%%%%%%%%%%
The properties of symmetric nuclear matter at $\rho_{0}$ is presented in Table~\ref{tab:matter-properties}.
It is found that the effect of gluon and pion decreases $M_{N}^{\ast}$, while it enlarges $K_{0}$.
We also show the third-order incompressibility, $J_{0}$, the slope and curvature parameters of the nuclear symmetry energy, $L$ and $K_{\rm sym}$, and the 2nd derivative of the isobaric incompressibility coefficient given by $K_{{\rm sat},2}=K_{\rm asy}-\frac{J_{0}}{K_{0}}L$ with $K_{\rm asy}=K_{\rm sym}-6L$.
The values of $K_{0}$ in this study are a little bit larger than the empirical one, and this point would be considered in the future work.

\begin{figure}[b]
  \hspace*{-0.3cm}
  \includegraphics[width=8.2cm,keepaspectratio,clip]{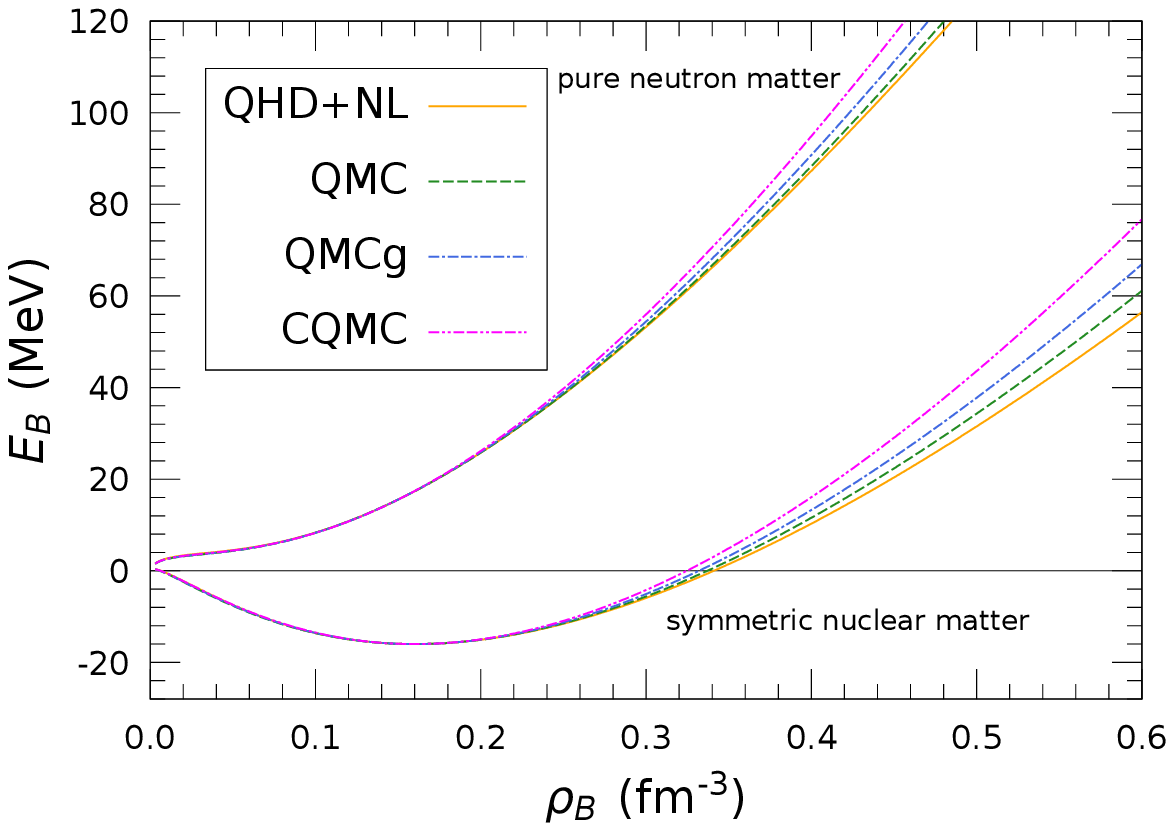}
  \hspace*{-0.5cm}
  \includegraphics[width=8.2cm,keepaspectratio,clip]{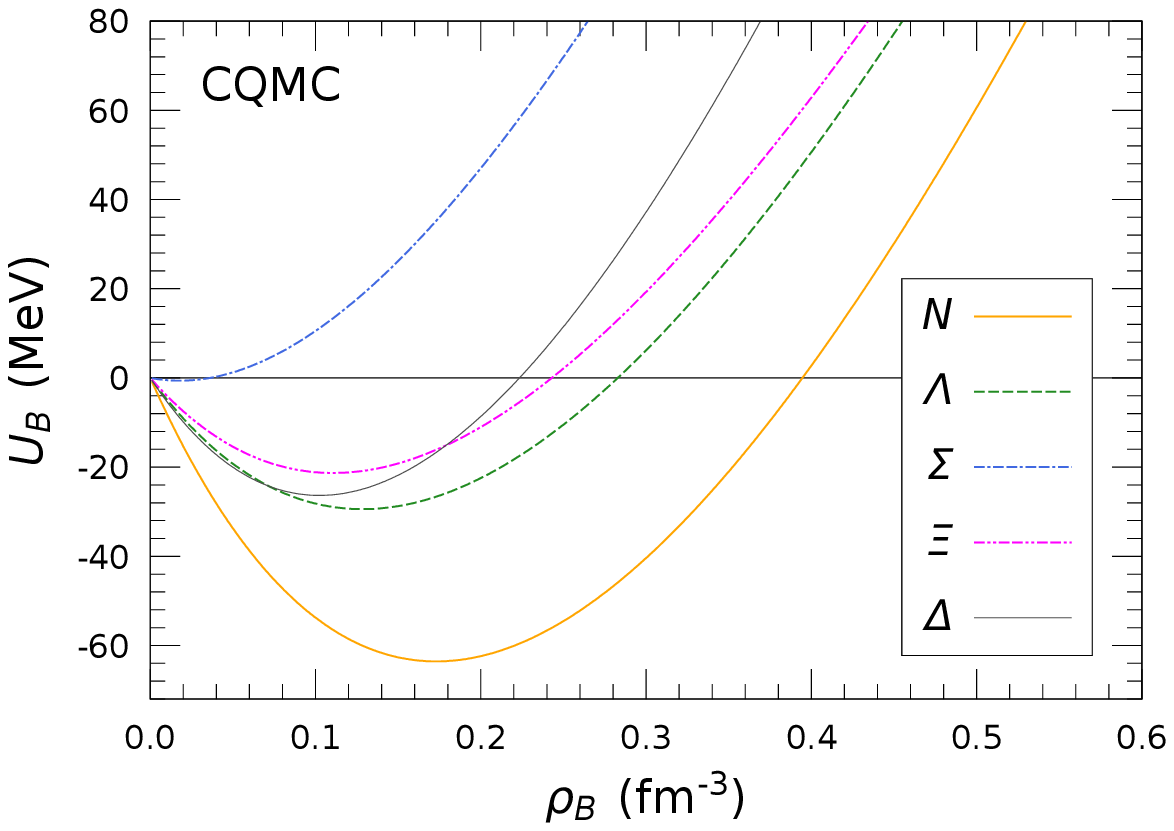}
  \caption{Nuclear binding energy per nucleon in symmetric nuclear or pure neutron matter (left panel) and potentials for the baryon, $B=N,\Lambda,\Sigma,\Xi,\Delta$, in symmetric nuclear matter (right panel) as a function of the total baryon density, $\rho_{B}$.}
  \label{fig:matter}
\end{figure}
The nuclear binding energy per nucleon in symmetric nuclear and pure neutron matter is shown in the left panel of Fig.~\ref{fig:matter}.
We find that, in both cases, the nucleon structure due to quarks and the hyperfine interaction due to gluon and pion increases the binding energy at densities above 0.3 fm$^{-3}$.
In the right panel of Fig.~\ref{fig:matter}, the baryon potentials in symmetric nuclear matter are presented.
Although the gluon and pion interactions between two quarks affect the difference among the potentials for $\Lambda$ and $\Xi$, it is a little difficult to reproduce the realistic potential for $\Sigma$, as already reported in Ref.~\cite{Miyatsu:2010zz}.
In the present study, we adjust the hyperon potentials so as to fit the following values suggested from the experimental
data of hypernuclei: $U_{\Lambda}(\rho_{0})=-28$ MeV, $U_{\Sigma}(\rho_{0})=+30$ MeV, and $U_{\Xi}(\rho_{0})=-18$ MeV~\cite{Miyatsu:2013yta}.

\begin{figure}[t]
  \includegraphics[width=8.0cm,keepaspectratio,clip]{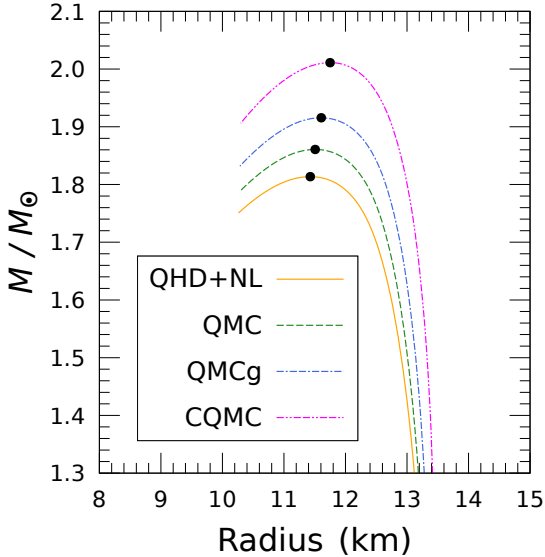}
  \caption{Mass-radius relations in the QHD+NL, QMC, QMCg, and CQMC models.
  The filled circle shows the maximum-mass point of a neutron star.}
  \label{fig:TOV}
\end{figure}
As for the properties of neutron stars, we show the mass-radius relations in Fig.~\ref{fig:TOV}.
Compared with the QHD+NL and QMC models, the maximum mass of a neutron star is pushed upwards because of the effect of quarks inside a baryon.
Additionally, the hyperfine interaction due to the gluon and pion exchanges between two quark enhances the maximum mass in the QMCg and CQMC model.
In particular, the pion-cloud effect is remarkably important to support the massive neutron stars.
In the present calculation, not only nucleons but also hyperons and delta-isobars are taken into account in the core of a neutron star.
In all cases, $\Lambda$, $\Xi^{-}$ and $\Xi^{0}$ are generated in the core, while the other particles do not appear in the density region below the maximum-mass point.

\section{Summary}

We have studied the effect of quark degrees of freedom inside a baryon and the hyperfine interaction due to the gluon and pion exchanges between two quarks on the properties of nuclear and neutron-star matter in SU(3) flavor symmetry.
It has been found that the variation of baryon substructure in matter plays an important role to support massive neutron stars with hyperons.
In particular, the pion-cloud effect is very vital to solve the $2M_{\odot}$ neutron-star problem.

\section*{Acknowledgments}

This work was supported by JSPS KAKENHI Grant Number JP17K14298.

\end{document}